\documentclass[conference,draftclsnofoot,onecolumn]{IEEEtran}%  
\usepackage[pdftex]{graphicx}
\usepackage[caption=false,font=footnotesize]{subfig}
\usepackage{multirow}
\usepackage{array}
\usepackage{subscript}
\usepackage{rotating}
\usepackage[nocompress]{cite}
\usepackage{titlesec}
\usepackage{enumitem}
\usepackage{flushend}

\usepackage{geometry}
\DeclareGraphicsExtensions{.pdf}
\newcolumntype{P}[1]{>{\centering\arraybackslash}p{#1}}

\if@titlepage
 \renewenvironment{abstract}{%
\titlepage
\null\vfil
\@beginparpenalty\@lowpenalty
\begin{center}%
\bfseries \abstractname
 \@endparpenalty\@M
 \end{center}%
 \par}%
 {\par\vfil\null\endtitlepage}
\else
 \renewenvironment{abstract}{%
 \begin{center}%
	\rmfamily\bfseries\fontsize{12}{14} \abstractname
 \end{center}%
 \par
 \itshape}%
\fi

\titleformat{\section}{\rmfamily\bfseries\fontsize{12}{14}}{\thesection.}{0.5em}%
		{}\relax
\titlespacing{\section}{0pt}{12pt}{12pt}

\titleformat{\subsection}{\rmfamily\bfseries\fontsize{11}{13}}{\thesubsection.}%
{0.5em}{}\relax
\titlespacing{\subsection}{0pt}{11pt}{11pt}

\titleformat{\subsubsection}[runin]{\rmfamily\bfseries\fontsize{10}{12}}%
{\thesubsubsection.}{0.5em}{}[.]
\titlespacing{\subsubsection}{0pt}{10pt}{0.5em}
\begin{document}

\title{A Graph Based Framework for Malicious Insider Threat Detection}
\author{\IEEEauthorblockN{Anagi Gamachchi, Li Sun and Serdar Bozta\c{s}}
\IEEEauthorblockA{School of Science, RMIT University, Melbourne,  Victoria, Australia\\
(anagi.gamachchi, li.sun, serdar.boztas)@rmit.edu.au}
}

\maketitle

This is a pre-publication version of the paper, which appeared in the Proceedings of the $50^{th}$ Hawaii International Conference on System Sciences (HICSS) 2017.

\begin{abstract}
While most security projects have focused on fending off attacks coming from outside the organizational boundaries, a real threat has arisen from the people who are inside those perimeter protections.
Insider threats have shown their power by hugely affecting national security, financial stability, and the privacy of many thousands of people. What is in the news is the tip of the iceberg, with much more going on under the radar, and some threats never being detected. We propose a hybrid framework based on graphical analysis and anomaly detection approaches, to combat this severe cyber security threat. Our framework analyzes heterogeneous data in isolating possible malicious users hiding behind others. Empirical results reveal this framework to be effective in distinguishing the majority of users who demonstrate typical behavior from the minority of users who show suspicious behavior.
\end{abstract}

\section{Introduction}
The battle between malicious but trusted insiders and organizations in safeguarding information assets is the biggest and fastest growing cyber security threat in this digital age. Insider threat, or the threat from a malicious insider who is defined as ``a current or former employee, contractor or business partner who has, or had, authorized access to an organization's network, system or data and intentionally exceeded or misused that access in a manner that negatively affected the confidentiality, integrity or availability of the organization's information or information systems" \cite{CommonSenseGuide3} has been identified as a primary concern within the cybersecurity community. Financial loss and reputation damage caused by this ``known unknown" cybersecurity threat far outweighs that caused by external attacks. Thus, the majority of private and governmental organizations have identified the severity of the threat posed by insiders and focused on security control improvements. However, the seriousness of this problem is still rising at an alarming rate threatening most critical infrastructure segments.  

One of the most recent articles from CSO magazine \cite{CSOstaff2016} compared the cost between external and internal attacks and noted that while it takes about 50 days to fix a data breach caused by an internal attack, it only takes 2 to 5 days in the case of external attacks. Moreover, ``attacks by malicious insiders are also the costliest to fix (\$145,000), followed by denial of service (\$127,000) and Web-based attacks (\$96,000)", indicating the severity of this problem. 
 
The unpredictable nature of human behavior makes this complex issue much more complicated than expected. This is aggravated by the mobility and hyper-connectivity of people. Insider threat research and surveys suggest this problem cannot be considered only as a data driven problem; it needs to be considered as data and behavior driven problem \cite{CappelliTheCERT2012}. A close examination of user behavior can spot trends and such information can be used in tightening radars on suspicious users. Different parameters govern users' day to day actions, and behavioral changes exposed in workplace environments will extend the possibilities of isolating suspicious users from the rest of the employees. Organizations can suffer after effects such as unmotivated employees, inefficient work behavior, if an innocent user is classified as suspicious. Thus, the decision of naming a person as suspicious should be a smooth but complex process. Obviously, insider threat detection will focus on isolating suspicious users from the others; but it may not be practical to point an employee as a malicious attacker. Also, the effectiveness of the process totally depends on the ability of analysis of many parameters as possible. Consideration of above facts led us to think of an insider threat detection framework as described in the rest of the paper.\\ 
\noindent
\textbf{\textit{Our Contribution}}:
In this paper, we propose a framework for isolation of malicious users based on graphical and anomaly detection techniques. The proposed architecture is given in Figure 1, and has two major components ``Graphical Processing Unit" (GPU) and ``Anomaly Detection Unit" (ADU). Data from multidimensional sources of an enterprise network is formatted and fed into the GPU, which generates a graph which represents interrelationships between informational assets of the network. These input streams can be from different, i.e., heterogeneous, informational sources with different data formats. These data streams can be from event logs (logon/logoff), email logs, HTTP records, social network access data and various HR records such as psychometric data. Once the informational assets are mapped into a network, several graph parameters are calculated for each user. Since the final goal %of this work 
is to isolate the most anomalous users from the rest, all the attributes are computed for individual users. The next task of the GPU is to generate induced subgraphs of each user for different levels of neighborhoods. Several relevant subgraph properties (vertex count, edge count, density, diameter and number of peers) are calculated for each level of subgraphs. Calculated graph and subgraph parameters are fed into the ADU. In parallel to the above process, time-varying data also fed into the ADU. The isolation forest algorithm is executed for isolating anomalous users within the ADU unit. Anomaly scores for each user is generated as the output of the ADU. These values are used in identifying and separating possible malicious users from the rest of the workforce.       

The remainder of this paper is organized  as follows. Section $2$ describes mostly related work contributed to insider threat detection research. Section $3$ is the dataset we used in this research while section $4$ described the adopted methodology. Experimental results are discussed in section $5$. Section $6$ concludes the paper indicating conclusions and future directions.

\begin{figure}[h]
\centering
\includegraphics[width=0.9\linewidth]{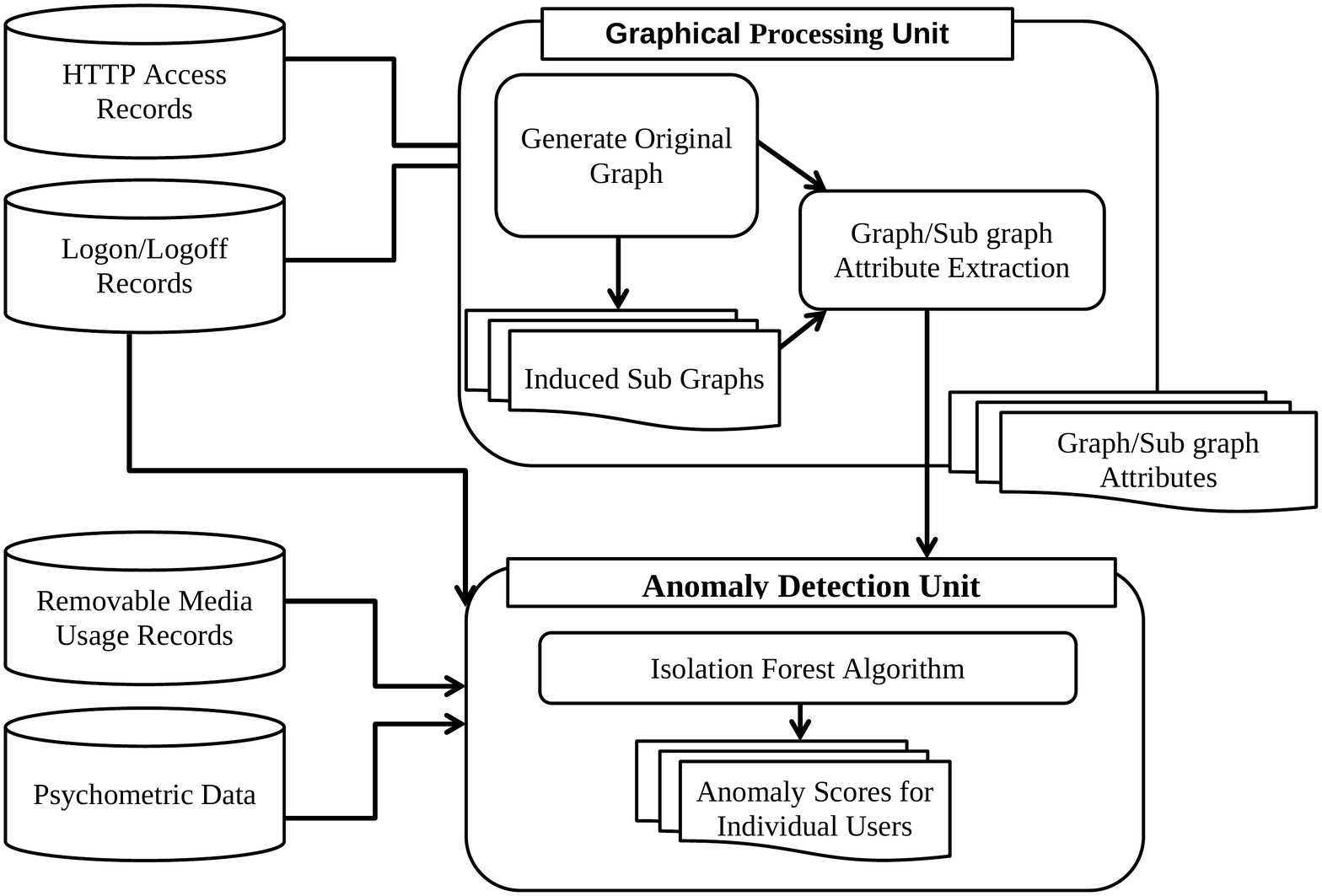}
\caption{Framework}
\label{figure 1}
\vspace{-10pt}
\end{figure}

\section{Related Work}
Intrusion Detection System (IDS) based approaches, visualization strategies, honeypot/honeynet approaches and system call based methods are several techniques adopted from external threat detection in finding solutions for insider problem \cite{Zeadally2012}. The work presented here is focused on the combination of graph-based techniques and anomaly detection approaches. This section will only discuss the most related work under the above two techniques specifically in the insider threat domain.

The specialized network anomaly detection (SNAD) model proposed by \cite{Chen2012} is a graph based approach used for detecting insider actions in collaborative information systems. In their model access logs are mapped into a bipartite graph. The similarity of users are compared based on the number of subjects a user accesses from a collaborative information system using the cosine similarity measure. In order to determine if a particular access is anomalous or not, the authors considered the influence of a user on the similarity of the access network by suppressing each user at a time. Even though they achieved better performance over their competitors (spectral anomaly detection model), they identified difficulties in implementation on real world networks.

Another study \cite{Brdiczka2012} proposed a proactive insider threat detection by graph learning and psychological modeling of users. The proposed model is a combination of structural anomaly detection and psychological profiling and explored the possibility of including dynamical properties of nodal attributes. They have evaluated results based on a publicly available gaming data set which might not be very similar to enterprise system and network data. Althebyn and Panda \cite{Althebyan2007} have also suggested the use of graph theory to formalize two components, knowledge graph and object dependency graphs. A knowledge graph represents knowledge units for a given insider and they are updated over the time. Dependency graph is a global hierarchical graph that shows all dependencies among various objects. Even though this model tries to include accumulated knowledge of the insider over time on systems and objects it can be improved by including several other parameters such as user's behavioral patterns and psychological aspects.

Another study performed by Nance and Marty \cite{Nance2011} introduced the use of bipartite graphs for identifying and visualizing insider threat. They tried to establish acceptable insider behavior patterns based on workgroup role classifications. Although this approach is quite useful for detecting certain insider threats, it has the limitation of a high false positive rate. The framework suggested by \cite{Eberle2009} is another graph-based approach for malicious insider threat detection which uses the minimum description length principle for detecting anomalous activities.
A recent paper \cite{Kent2015} proposed the use of authentication subgraphs for analyzing users behavior within an enterprise network utilizing a set of subgraph attributes in user profiling. Time series analysis of subgraphs and use of bipartite graphs are also introduced in their work, targeting on a much comprehensive analysis in their ongoing work. Another related framework, the BANDIT (Behavioral Anomaly Detection for Insider Threat) \cite{Berk} also proposed a two-stage anomaly detection approach in managing insider attacks which compared user's behaviour based on own and peer baselines.   

\section{The Data Set}

Due to the lack of availability of proper insider threat datasets we have utilized the insider threat dataset published by CERT Carnegie Mellon University for this research \cite{CERTDataset}. The dataset \textit{``R4.2.tar.bz"} has been used for this analysis. According to the dataset owners, this is a ``dense needle" dataset with a fair amount of red team scenarios. This dataset consists of six broad types of data records (HTTP, logon, device, file, email and psychometric) of 1000 employees over a 17 months period. All HTTP records contain user, PC, URL and web page content with time stamps. ``Logon.csv" consists of user logon/logoff activities with the corresponding PC with timestamps. ``Logon" and ``Logoff" are the two types of activities can be found in data records. ``Logon" activity corresponds to either a user login event or a screen unlock event, while the ``Logoff" event corresponds to user logoff event. Screen locks are not recorded in this dataset. The third data file ``device.csv" is a collection of data records of removable media usage. It indicates insert/remove actions with the relevant user, PC, and timestamp. Details of file copies are stored in ``file.csv" file with date, user, PC, filename, and content. To get the friendship network of users, the CERT dataset provides email communication records including from, to, cc and bcc fields. ``psychometric.csv" provides psychometric scores based on big five personality traits or five-factor model (FFM) for the definition of personality. 

Among the different work roles which in the dataset, we performed our analysis with three job roles. Table 1 shows the statistics of selected data records.

\renewcommand{\arraystretch}{1.2} 
\begin{table*}[h]
\caption{Data statistics}
\centering
\begin{tabular}{|p{0.19\linewidth}|p{0.17\linewidth}|p{0.06\linewidth}|p{0.07\linewidth}|p{0.07\linewidth}|p{0.07\linewidth}|p{0.07\linewidth}|p{0.08\linewidth}|}
\hline Functional Unit & Department & Number of Users & HTTP  & Logon & Device & File & Psychometric\\
\hline  Research And Engineering & Engineering & 129 & 4,196,817 & 101,782 & 67,916 & 75,335 & 129 \\
\hline  Research And Engineering & Software Management & 101 & 3,295,774	& 82,187  & 44,049 & 58,173 & 101 \\
\hline  Research And Engineering & Research & 101 & 3,332,576 & 79,362 & 30,906  & 41,292 & 101 \\
\hline 
\end{tabular}
\vspace{-5pt}
\end{table*}

\section{Methodology}
The goal of this paper is to introduce a framework for mitigating the insider threat problem using a combination of graph-based approach and an anomaly detection technique. This framework will utilize multidimensional inputs such as user interactions with hardware assets, web access records, email correspondences and psychometric figures. The graph-based approach is a prominent method of identifying inter-relationships between multidimensional entities. A user's interactions with devices are illustrated in a weighted, undirected large scale bipartite graph $G=(V,E,W)$, where $V$ is the set of vertices (users), $E$ is the set of edges, and $W$ is the set of edge weights. Set of vertices comprises of two types of entities, users and devices while edges represent user's interaction with the device. Edge weights correspond to the number of ``Logoff" activities which appeared during the whole time duration of the dataset between an individual user and a device. Graph visualization was carried out using NodeXL \cite{NodeXL} and all the other calculations were done using the R statistical computing language \cite{R}. The following subsections describe the theoretical background and the implemented methodology in detail.

Even though the dataset comprises of both ``logon" and ``logoff" records for individual users, we utilize only the ``logoff" events for network mapping. The reason behind this is that we cannot distinguish logon activities, and screen unlocks as they both recorded as ``logon" events. However, the screen locks are not recorded, and only the logoff events are recorded as ``logoff" events. The representation of the use of removable media also can be represented by an edge between the corresponding user and the device. However, this will convert the graph into a multigraph where the existence of multiple edges is possible among two vertices. To keep the simplicity of the first phase of analysis, we shall exclude the representation of such edges. Integration of the friendship network into the same network would change the structure of the network, and it will lose the multipartite property. Therefore, inter-user relationships based on email data has not been considered for graphical representation. The following graph attributes for individual users are captured from the above graph $G$ for further analysis. 

\subsection{Graph  Parameters}

\textbf{\textit{User's vertex degree $(d_{u})$}}:
The degree of a vertex is the number of edges connected to it. In the context of this analysis, we calculated the degree only for the users. Therefore this value represents the number of devices accessed by an individual user.

\subsection{User Subgraph (USG) Parameters}

\begin{figure}[h]
\centering
\includegraphics[width=0.5\linewidth]{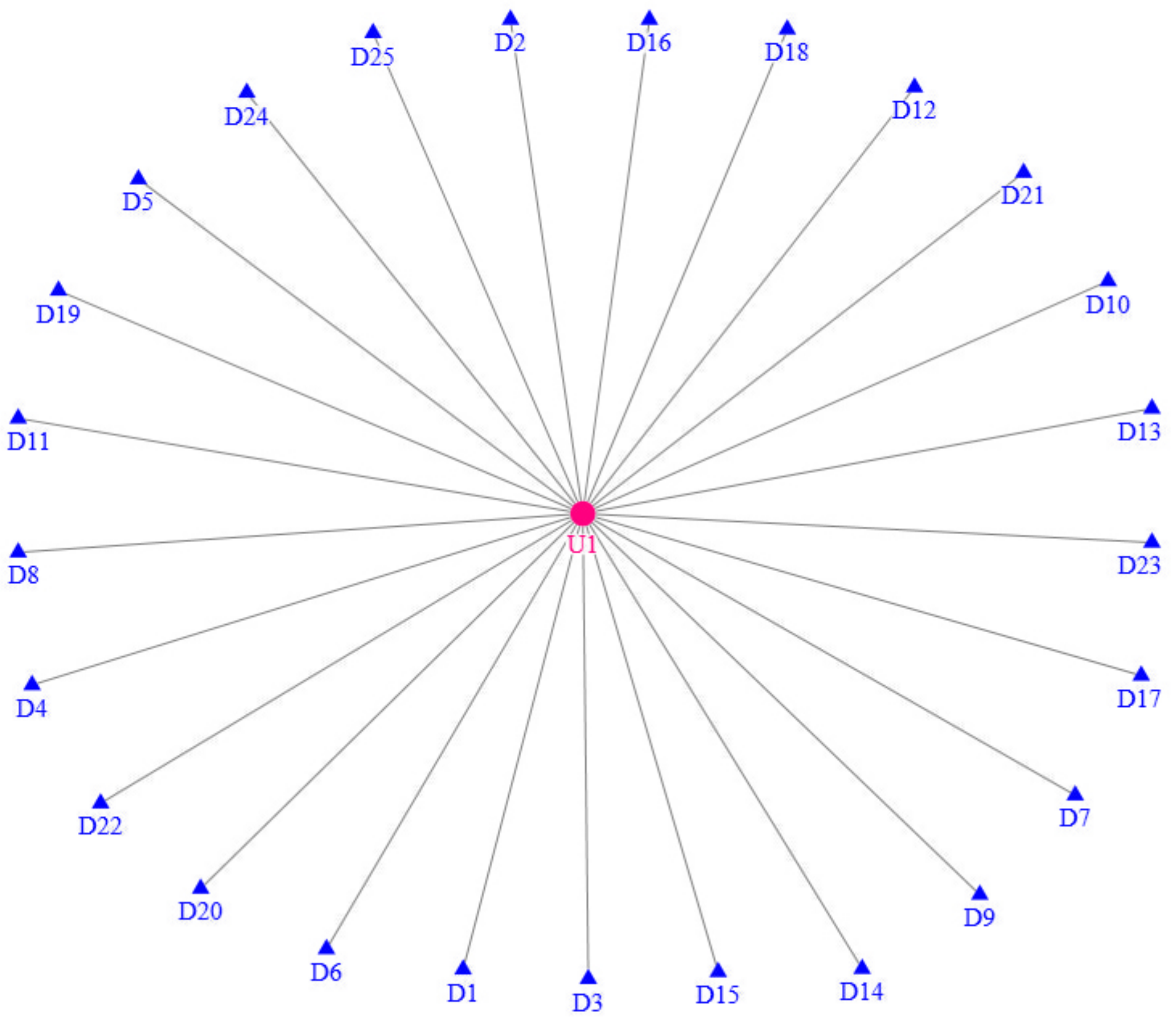}
\caption{First order subgraph topology for all users}
\label{figure 2}
\vspace{-10pt}
\end{figure}

As the next step of the analysis, the focus is on the construction of subgraphs for each user. A deep dive into the different order of subgraphs is how we extract relevant graph parameters in graph analysis. In this work user subgraphs are constructed until the fifth order of neighborhood. We define the user sub-graph (USG) as a weighted undirected graph $G_{u}=(V_{u},E_{u},W_{u})$ for the entire period of the dataset. $V_{u}$ represents the vertex set of $G_{u}$ while $E_{u}$ represents the edge set of $G_{u}$. Link weights are similar to that of the original graph. We have noticed that all the first order subgraphs have a star topology as shown in Figure 2. The number of external nodes can be within $(1:\ $total number of devices$)$. This factor is because of the disjoint nature of the two type of vertices. Even though we cannot extract much information on device access similarities with first order subgraphs, we will continue to use following subgraph properties of first order subgraphs in addition to the higher order subgraph properties for the completeness of this analysis.

\begin{enumerate}[leftmargin=*]
	\item Vertex count $v_{u_{j}}$ for $j=1:5$
	\item Edge count $e_{u_{j}}$ for $j=1:5$
	\item Density $p_{u_{j}}$ for $j=1:5$
	\\ The density of a graph is the ratio of edges to all possible edges given the number of vertices.
	\item Diameter $d_{u_{j}}$ for $j=1:5$
	\newline{The diameter of a graph is the largest shortest path between any two vertices.}
	\item Number of peers $p_{u_{j}}$ for $j=1:5$
	\\Since the main focus of this analysis is to identify the most anomalous users, we have chosen this parameter for evaluation in addition to the basic graph properties.
	\end{enumerate}
	
The distribution of the above properties has been illustrated as histograms and further discussed in the ``Experimental Results" section of this paper.

\subsection{Time Dependent Parameters}
We believe that it is not sufficient to consider only the above parameters in identifying malicious insiders without the temporal properties. To carry out a  complete and comprehensive analysis, the following time-based parameters and personality values have been  identified as important input parameters for the anomaly detection algorithm.

\subsubsection{Individual logon logoff events}
This parameter can be used in identifying users abnormal logon/logoff activities as most disgruntled insiders tend to commit malicious activities after hours \cite{CappelliTheCERT2012}. Identifying users' baseline behavior on system/device access is  an essential part of malicious insider threat detection problem. For each user, four parameters (minimum, maximum, mean and mode) logon and logoff values have been calculated. Those four parameters are also fed as an input parameters to the anomaly detection unit.
	
\subsubsection{Removable media usage events}
Removable media is among the most popular method used in theft of Intellectual Property (IP) in extracting confidential information from organizations \cite{CappelliTheCERT2012}. Tracking the use of removable media can  be an excellent information source for identifying suspicious events by trusted insiders . Baseline behavior of removable media usage is captured by the minimum, maximum, mean and mode time of ``Insert" and ``Remove" activities as in the logon/logoff event analysis. Time gap between consecutive ``Insert - Remove" action has also been identified as a good source of information to capture large file downloads.  The daily number of files copied by an individual is also used in this analysis.
	
\subsubsection{Web Access Patterns}
We can think of users' online behavior as a reflection of their offline behavior, as they tend to publish their feelings, thoughts, likes and dislikes through social media. In addition to the above fact, web access patterns is also a good indication of their online behavior. Disgruntled insiders tend to access competitors websites and recruitment agency websites to understand and gather information on potential opportunities. We have identified that the users online behavior analysis will be comprehensive if we include multiple social media data sources and web access records. However due to the limitations of data availability on all these domains we will be strict only to web access records for this research work. Also, it is evident that the content of web pages may have a direct link with users suspicious behavior. However, we will again restrict content analysis of web pages in this analysis and will consider it as a future direction of our continuing work.

\subsection{Personality Parameters}
\textbf{\textit{Psychometric data }}:
Psychological behavior is one of the other aspects linked to insider attacks. Sudden behavioral changes can be indications of misuse of privileges. Verbal behavior, personality traits, unauthorized absences, aggressive behavior, are a few indicators which can be considered as small markers which come before the big attack. Therefore nontechnical behavioral changes are captured through the psychometric data provided in the dataset.
	
Table 2 is a summary of all properties we have identified in this analysis to consider as input parameters in to the isolation forest algorithm to separate most anomalous users. Due to space limitations, we have listed some pairs of parameters in the same line in the table, e.g., the line "Minimum/Maximum Insert Time" covers two parameters,  "Minimum Insert Time", and "Maximum Insert Time".

\renewcommand{\arraystretch}{1.2} 
\begin{table}[!t]
\caption{Selected parameter set}
\begin{center}
\begin{tabular}{|p{0.32\linewidth}|p{0.55\linewidth}|}
\hline \textbf{Module} & \textbf{Parameter}\\
\hline Graph & Degree of vertex\\
\hline \multirow{4}{*}{Sub Graphs} & Vertex Count\\
\cline{2-2}  & Edge Count\\
\cline{2-2}  & Density\\
\cline{2-2}  & Weighted Diameter\\
\cline{2-2}  & Number of Peers\\
\hline 
\multirow{4}{0.32\linewidth}{Logon/Logoff Events} & Minimum/Maximum Logon Time\\
%\cline{2-2}  & Maximum Logon Time\\
\cline{2-2}  & Mean/Mode Logon Time\\
%\cline{2-2}  & Mode Logon Time\\
\cline{2-2}  & Minimum/Maximum Logoff Time\\
%\cline{2-2}  & Maximum Logoff Time\\
\cline{2-2}  & Mean/Mode Logoff Time\\
%\cline{2-2}  & Mode Logoff Time\\
\hline \multirow{6}{*}{Removable Media} & Minimum/Maximum Insert Time\\
%\cline{2-2}  & Maximum Insert Time\\
\cline{2-2}  & Mean/Mode Insert Time\\
%\cline{2-2}  & Mode Insert Time\\
\cline{2-2}  & Minimum/Maximum Remove Time\\
%\cline{2-2}  & Maximum Remove Time\\
\cline{2-2}  & Mean/Mode Remove Time\\
%\cline{2-2}  & Mode Remove Time\\
\cline{2-2}  & Maximum number of daily file copies\\
\cline{2-2}  & Mode of number of daily file copies\\
\hline Web Access Patterns & Number of Unique URLs \\
\hline \multirow{5}{0.32\linewidth}{Psychometric Observations} & O (Openness to experience) \\
\cline{2-2}  & C (Conscientiousness)\\
\cline{2-2}  & E (Extroversion)\\
\cline{2-2}  & A (Agreeableness)\\
\cline{2-2}  & N (Neuroticism)\\
\hline 
\end{tabular}
\label{Table 2}
\end{center}
\vspace{-20pt}
\end{table}

\subsection{Anomaly Detection}
Due to the complex nature of insider threat problem, it is extremely hard to pinpoint a user as a malicious insider. Therefore, the first step should be the identification of possible malicious insiders who are maximally deviating from peers as well as their normal behavior. Therefore, as the second stage of our analysis, we will focus on implementing an anomaly detection algorithm based on the important graphical properties and time dependent properties identified at the previous stage of this analysis. The anomaly detection algorithm adopted in this analysis is the ``Isolation forest" algorithm, which stands out in effectively separating anomalous events from the rest of the instances \cite{IForest}. 

\textbf{\textit{Isolation Forest Algorithm - iForest}}:
The isolation forest algorithm is a model-based approach which explicitly isolates anomalies without constructing a typical profile instance. Linear time complexity with a low constant and low memory requirements drives us to use it in our experiments as the enormous amount of information need to be analyzed in the field of insider threat. The use of the isolation forest algorithm for this work is part of the overall research effort within our research group  at RMIT University and CA Pacific, with the details as presented in \cite{Li2016}, where it is applied to a very large enterprise system for anomaly detection. 

This algorithm also performs well with a large number of irrelevant attributes and instances where training data set does not contain any anomalies. This method generates an ensemble of iTrees for a given dataset and the instances with the short average path of iTrees are considered to be anomalies. If the calculated anomaly score value, $s$ is very close to $1$ it can be regarded as a definite anomaly. Instances with $s$ much smaller than $0.5$ can be considered normal situations. If all the instances return $s \approx(0.5)$, then the entire sample deemed to be not having any distinct anomalies.

Based on the above-described algorithm, anomaly scores were calculated for each user, for each order of subgraph for (1:5) separately, based on the five graph properties identified in subsection 4.2. In this case the \textit{iForest} algorithm is executed considering $5$ input parameters. We believe it would be much effective if we incorporate parameter values calculated for the different order of subgraphs when calculating anomaly scores. Therefore anomaly scores corresponding to subgraph properties have been computed using $25$ distinct values obtained for 5 various parameters of $1^{st}, 2^{nd}, 3^{rd}, 4^{th} and\ 5^{th}$ order of subgraphs. Similarly, anomaly scores correspond to graph parameters, time-dependent parameters and personality parameters (as summarized in Table 2) were calculated independently using iForest algorithm. Finally, anomaly scores correspond to each user is calculated as a combination of all the parameters described in Table 2, in which case the number of input parameters for the algorithm was 50. Breakdown of the number of parameters has been summarized in Table 3.  

\begin{table}[!t]
\caption{Summary of number of parameters}
\begin{tabular}{|p{0.45\linewidth}|P{0.40\linewidth}|}
\hline \textbf{Property} & \textbf{Number of Input parameters}\\
\hline Graph properties & 1\\
\hline Sub graph properties & 25\\
\hline Logon/Logoff behavior & 8\\
\hline Removable media usage & 10\\
\hline Web access patterns & 1\\
\hline Psychometric observations & 5\\
\hline
\end{tabular}
\label{Table III}
\vspace{-10pt}
\end{table}

\section{Experimental Results}
This section is dedicated to a comprehensive discussion of results obtained through our analysis. The discussion is based on the ``Research and Engineering/Engineering" work role, and final results for all three work roles considered in this analysis are summarized in Table 4.

Figure 3 is an illustration of the user's device access network. Users are represented in spheres while devices are represented by triangles. Vertex size corresponds to the degree of the vertex, which is an indication of the number of devices accessed by a particular user. The width of all edges corresponds to the number of ``Logoff" events which occurred during the entire period of the dataset. We continue to use the graphical representation of informational assets as in our previous work \cite{Anagi}, as it can be used to precisely indicate interrelationships between informational assets. Also, it is an efficient means of extracting basic essential parameters of massively dense log data. 

\begin{figure}[h]
\centering
\includegraphics[width=0.5\linewidth]{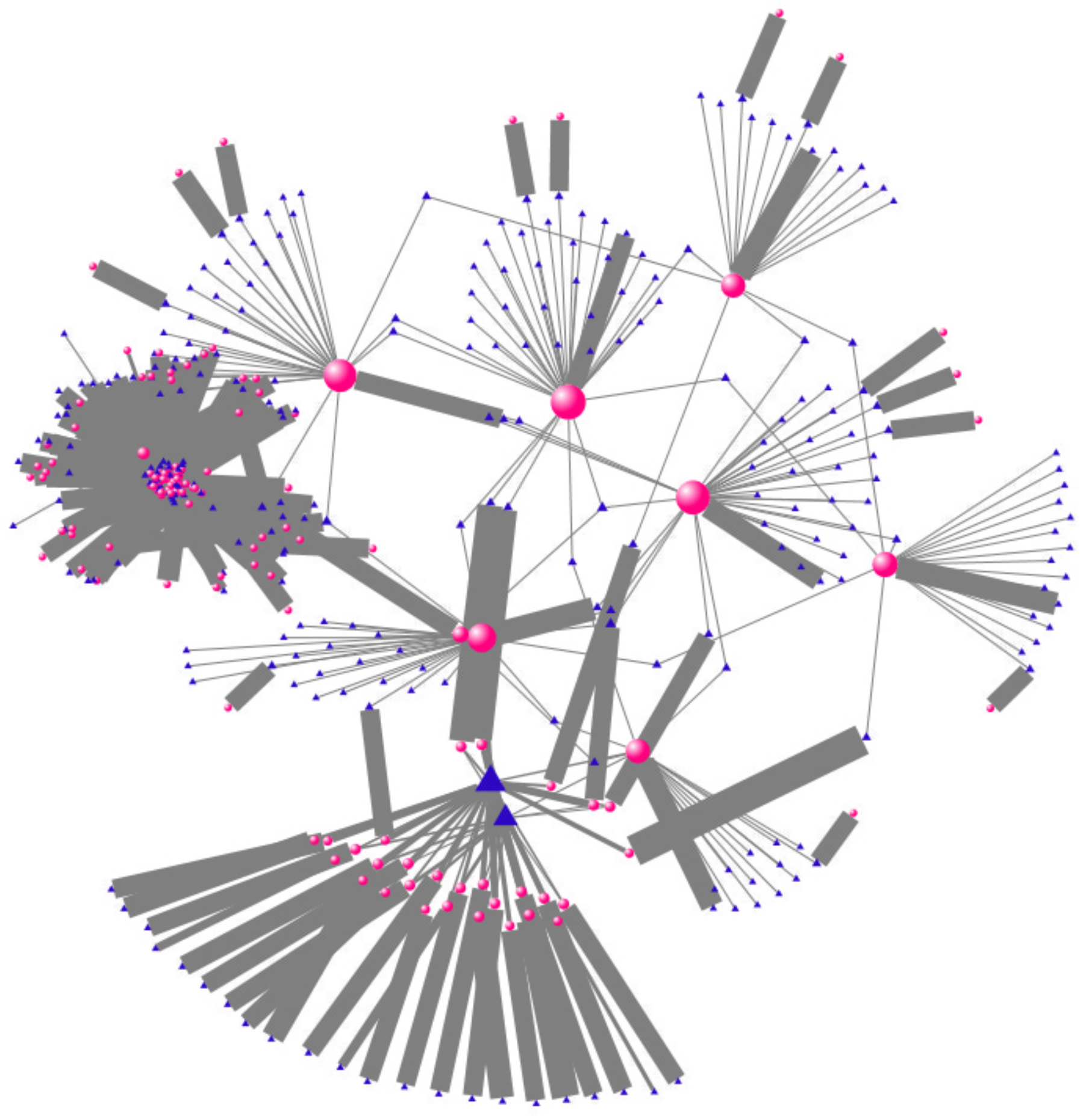}%
\caption{Users' device access network}
\label{figure 3}
\vspace{-5pt}
\end{figure} 

\subsection{Graph Parameters}
The individual degree distribution is illustrated as a histogram in Figure 4. The degree of a user is the number of devices a user access in this analysis. This figure reveals that the majority of users have a lower degree while the minority of users have the larger degree compared to others. We can think of two possible reasons behind the few number of users in the tail of the distribution. Either these employees are assigned to multiple devices to perform their day to day operations or an anomalous behavior. As previously mentioned, we should not directly conclude any of the insiders in the tail of the distribution as suspicious just by looking at the number of devices they accessed. However, we can think them as high-risk profiles among the others.    

\begin{figure}[!t]
\centering
\vspace{-10pt}
\includegraphics[width=0.5\linewidth]{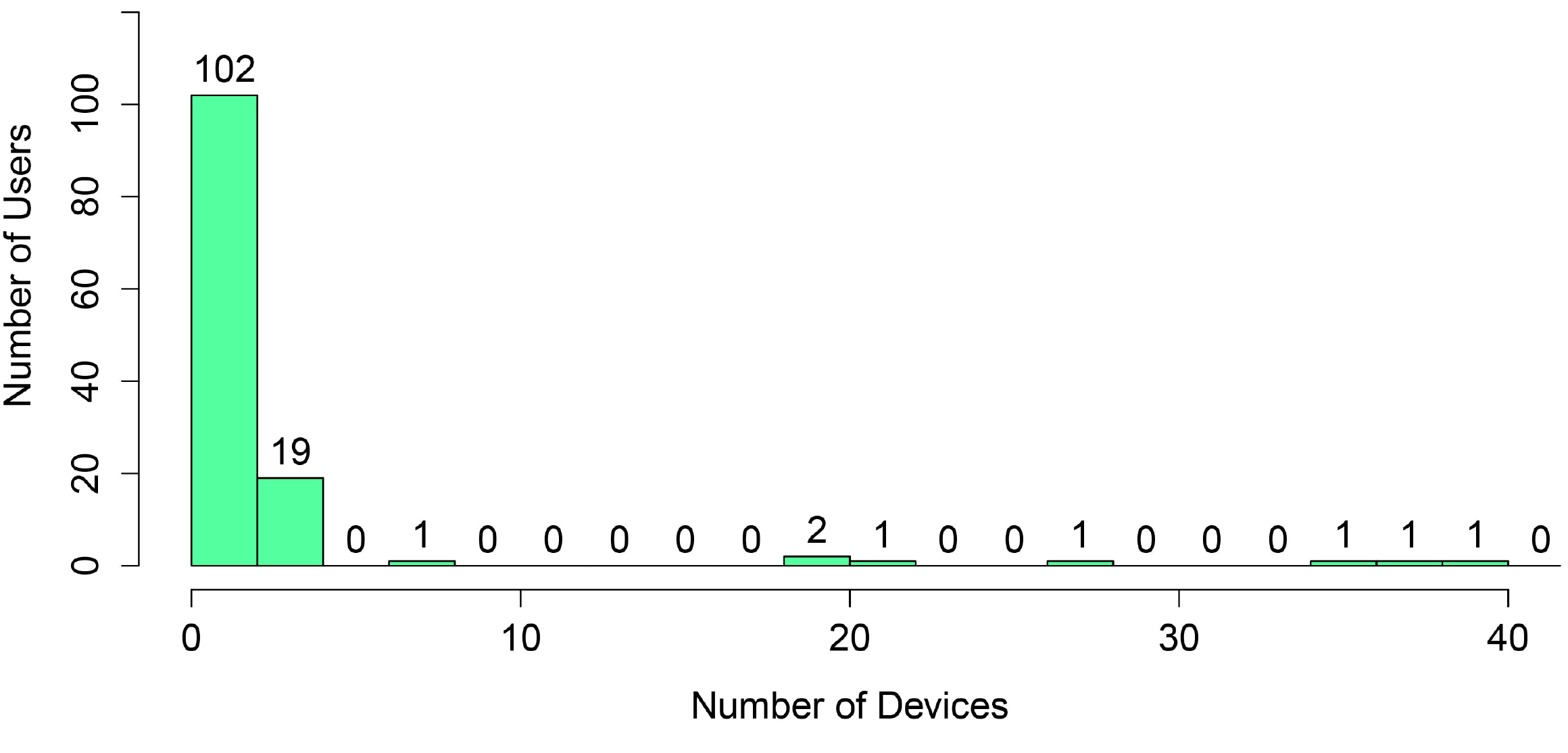}
\caption{Degree distribution of users}
\label{figure 4}
\vspace{-5pt}
\end{figure}

\subsection{Subgraph Parameters}
Figure 5(a) illustrates the histogram of vertex count for the different order of user subgraphs. These histograms show the majority of users have a small number of vertices in their subgraphs while the minority of users have a larger number of vertices in their subgraphs resulting much complex user subgraphs. Figure 5(b) is an illustration of the distribution of edge count across the different order of subgraphs. These values also follow a distribution which is very similar to vertex count. The subgraph density and the weighted diameter for all USGs also shown in figure 5(c) and figure 5(d) respectively. Density histograms show a similar pattern for most of the cases. Even though we could not find any obvious reason for this nature of distribution, we think that subgraph density is an important attribute for this kind of evaluations. Therefore, we continued to use that parameter as an input for the ADU. Diameter distribution for lower order subgraphs shows similar behavior with a single peak data bar while higher order subgraphs show similar behavior with two distinct peaks. These peaks are an indication of cluster/clusters of users who have similar behavioral patterns. The other subgraph property, number of peers also indicates two broader groups of users corresponds to two significant data bars and few other small groups of users. These results indicate the significance of higher order subgraph analysis in finding little clues among the enormous amount of data.
 
\begin{figure*}[h]
		\centering
		\subfloat[Vertex count]{\includegraphics[scale=0.20]{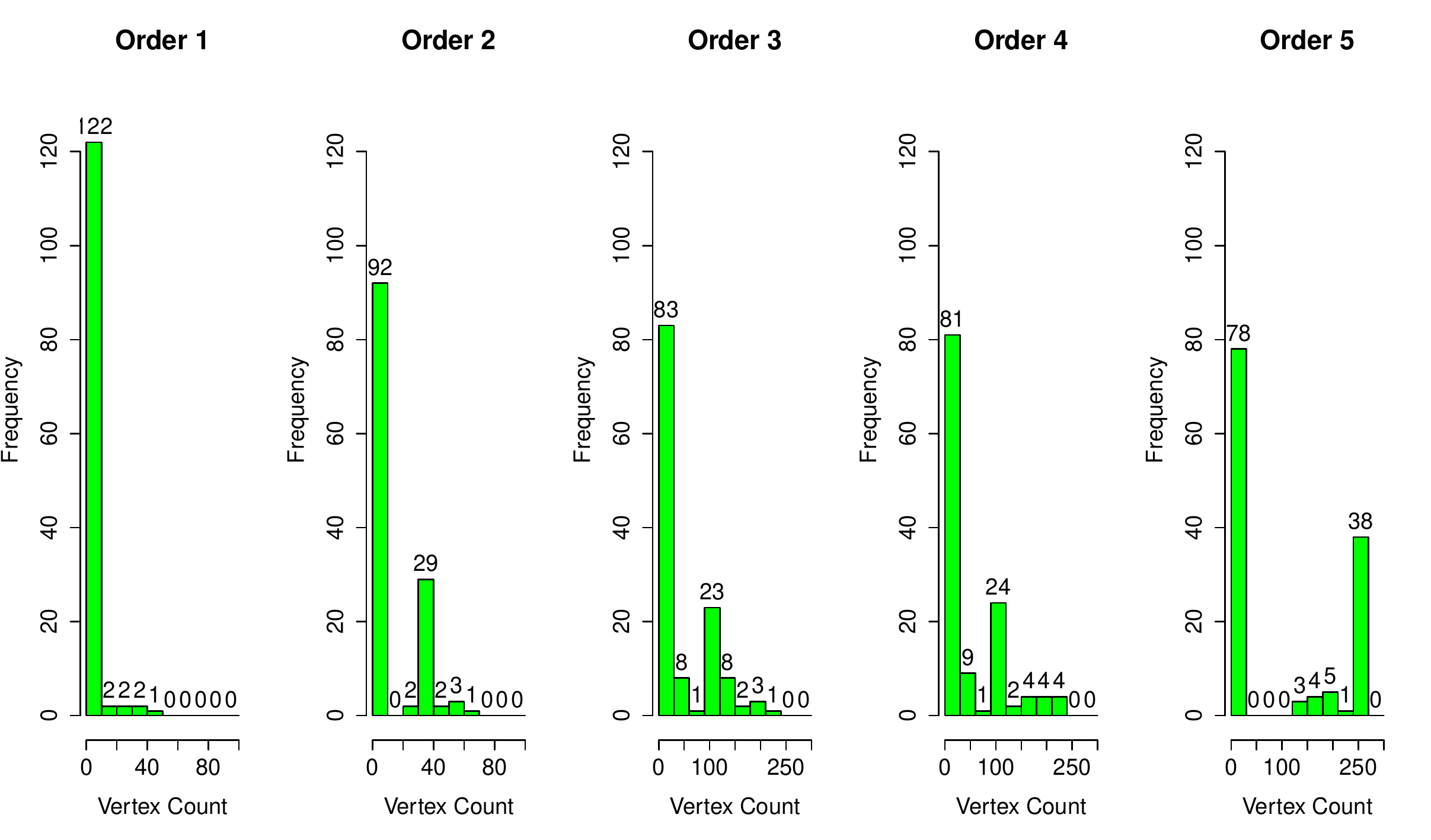} 
		\label{Vertex_count}}
		\subfloat[Edge count]{\includegraphics[scale=0.20]{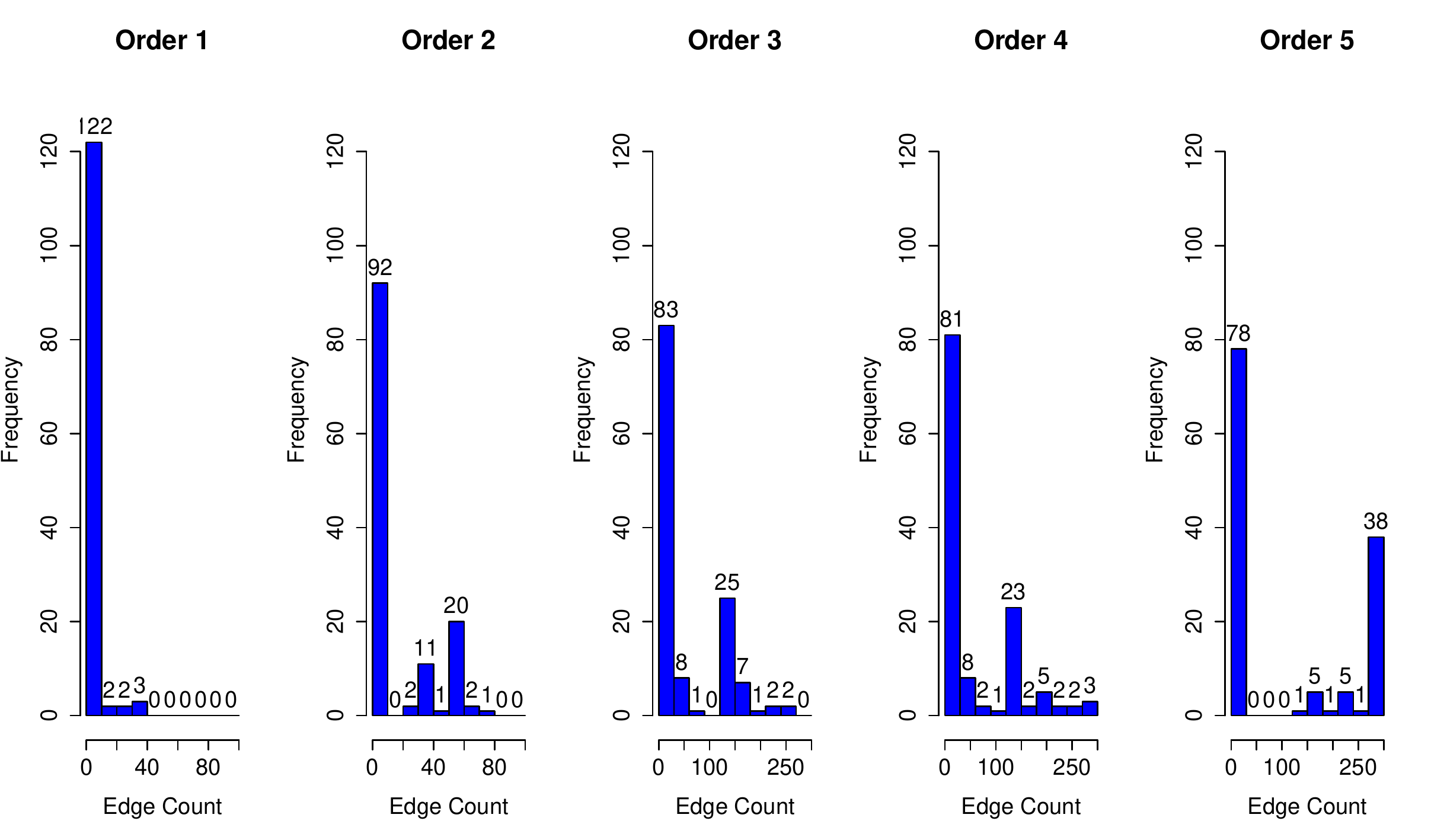}	
		\label{Edge_count}}
		\subfloat[Density]{\includegraphics[scale=0.20]{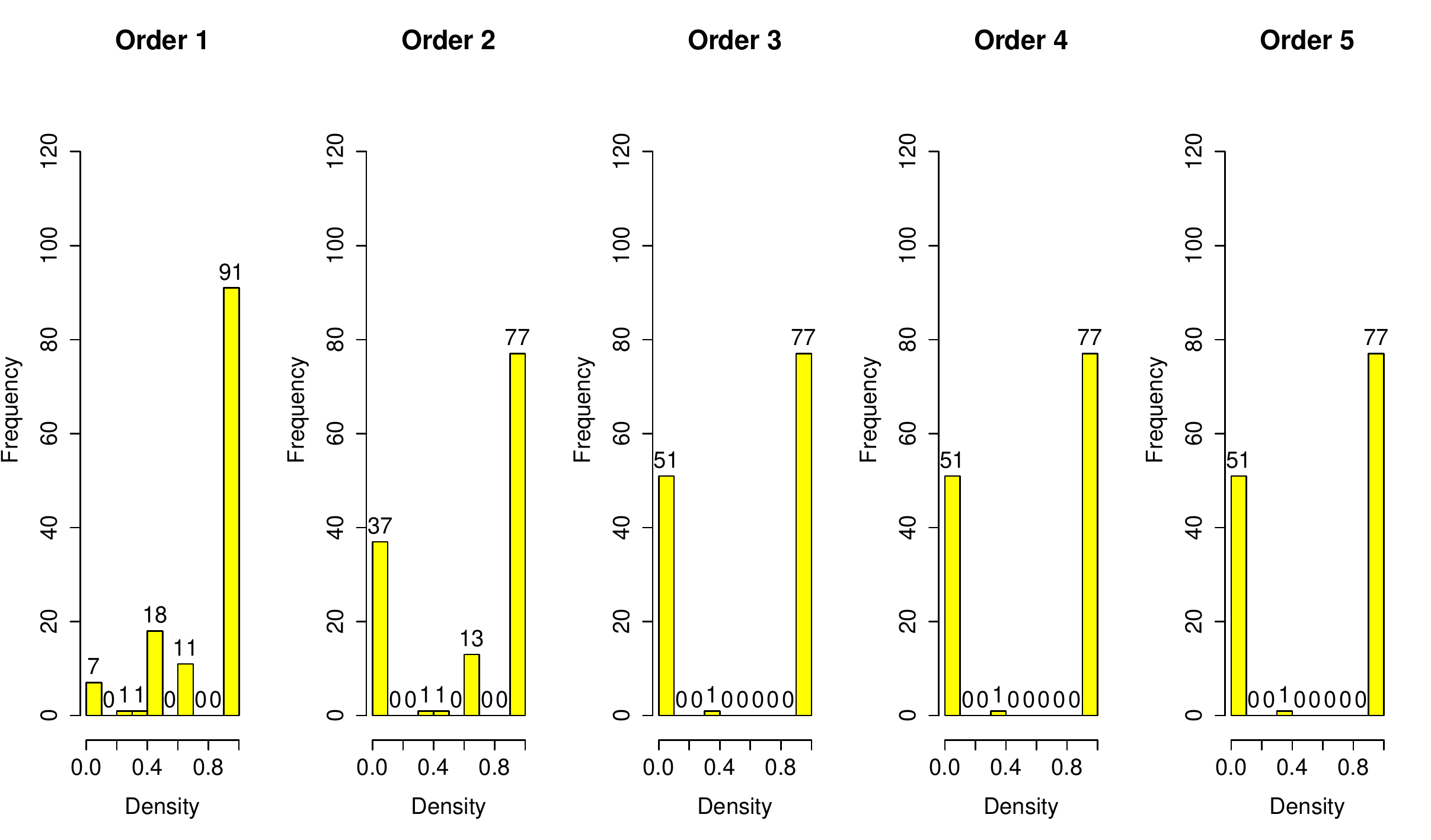}
		\label{Density}}
		\hfil
		\subfloat[Weighted diameter]{\includegraphics[scale=0.23]{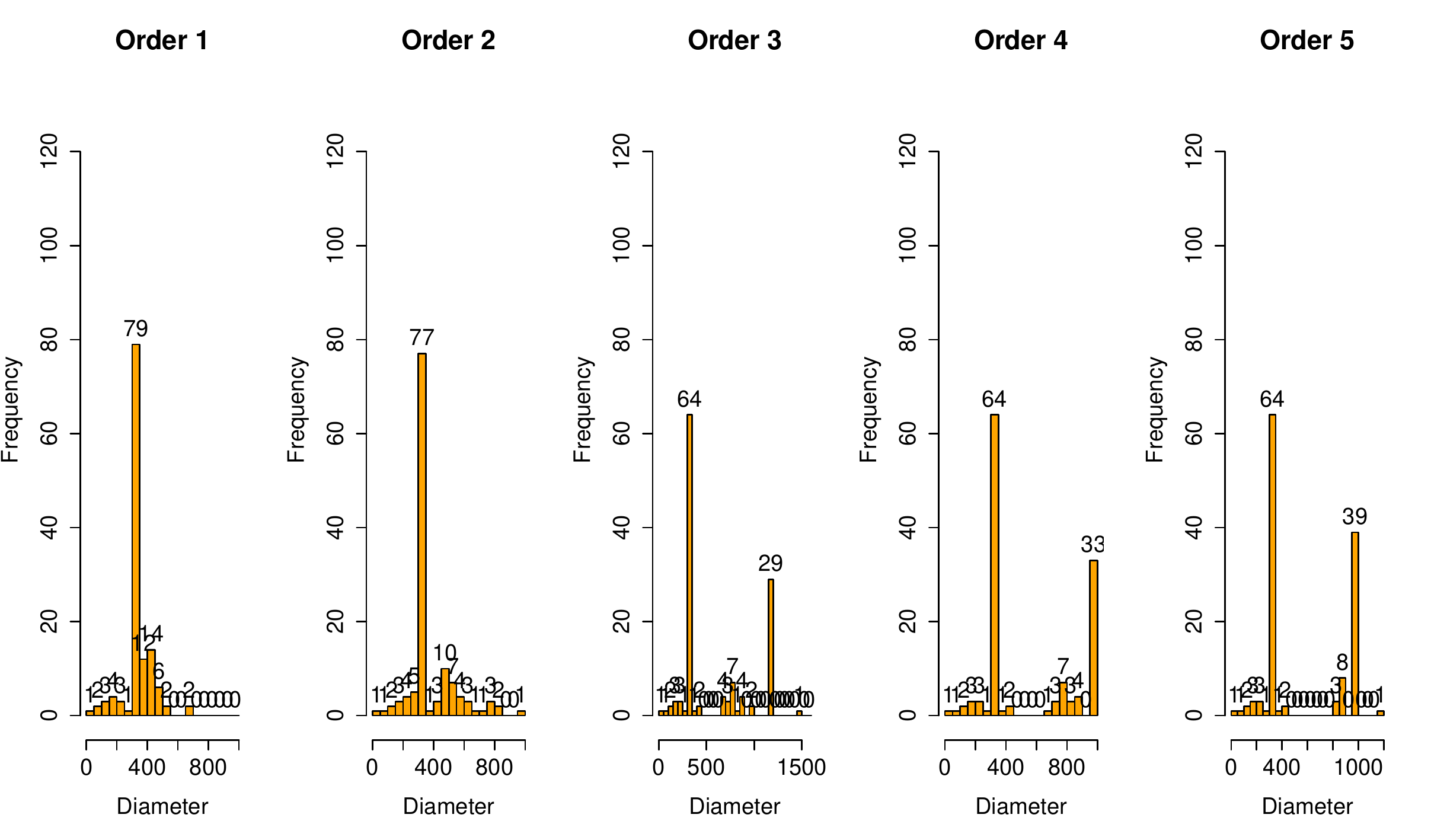}	
		\label{Weighted Diameter}}
		\subfloat[Number of peers]{\includegraphics[scale=0.23]{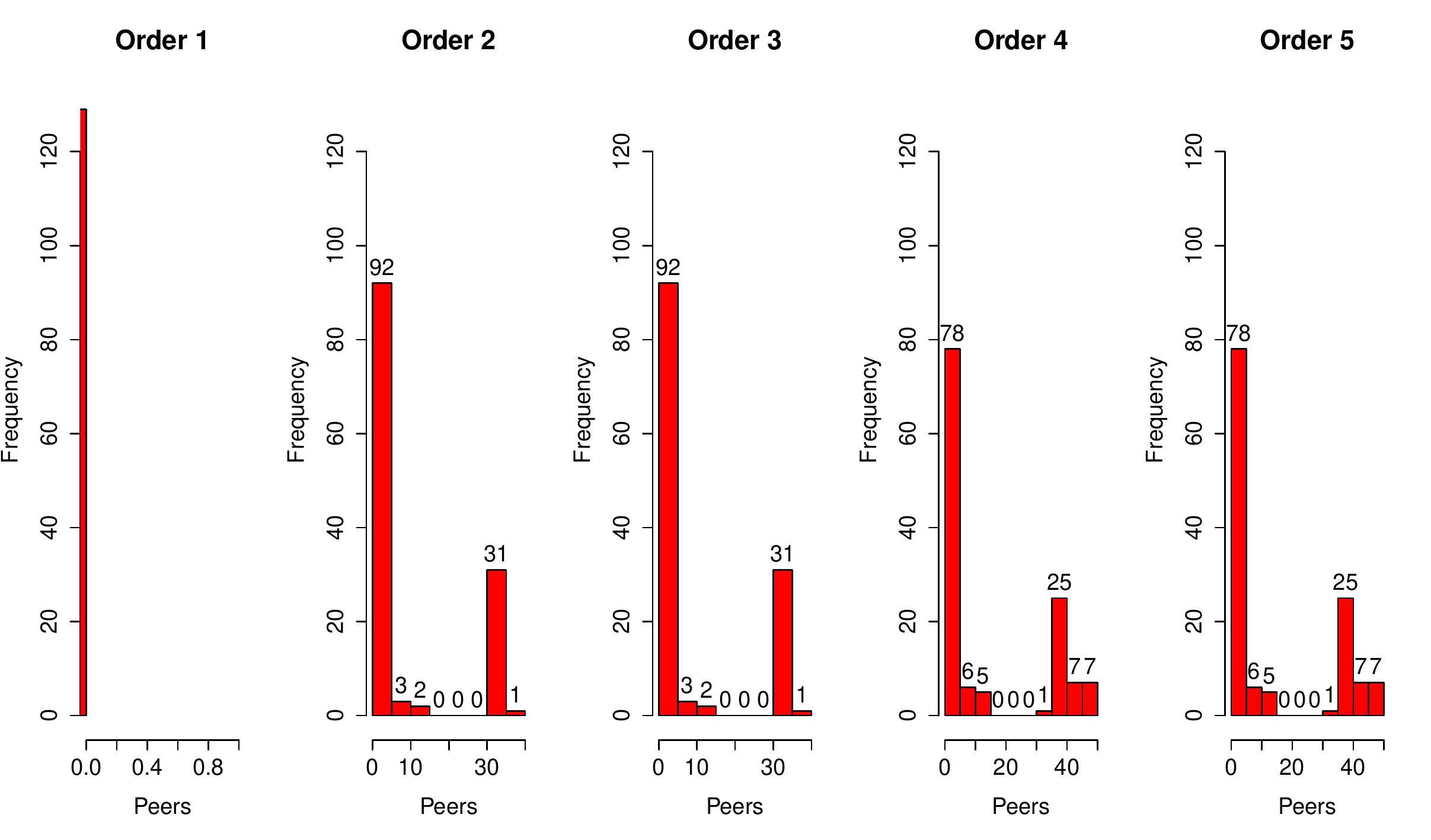}	
		\label{Number of Peers}}
		\caption{Histograms of subgraph properties}
		\label{figure 5}
		\vspace{-5pt}
\end{figure*}

\subsection{Time Dependent Parameters}
This subsection discusses the results obtained for other time-varying properties, which were identified as some of the other governing parameters of insider threat problem. 

\subsubsection{Individual Logon-Logoff Behavior}
Figure 6(a) is an illustration of users logon behavior for the entire period of the dataset. By looking at this graph, it is evident that the majority of logon activities occur during early office hours, which can be interpreted as the first logon event of the day. There are some other logon events, especially when we consider the ``maximum" logon time which occurs during regular working hours, which can be treated as logins followed by screen locks during the day. The logon times which we need to pay more attention are the events which happen during after office hours. We can identify a few users who have minimum and maximum logon times occurred during the late night, which might be unusual for normal operations. In real world enterprise networks, we can expect system user logon activities during this type of time periods for scheduled jobs such as backups, log rotations and routine activities. But if we find such logon activities for non-system users, that is for human users, it needs to be further investigated to differentiate between a genuine or a suspicious activity.\\
One of the other critical parameters of insider threat detection, the ``logoff" behavior of users are illustrated in Figure 6(b). This graph also shows the mean, mode, minimum and maximum logoff times of each user for the entire period. As can be seen on the graph majority of ``logoff" events happen during late office hours. As in the case of ``logon" behavior we are concern about after hours logoff events which are abnormal compared to the majority events happen during regular business hours.

\begin{figure*}[h]
		\centering
		\subfloat[Logon behavior]{\includegraphics[scale=0.30]{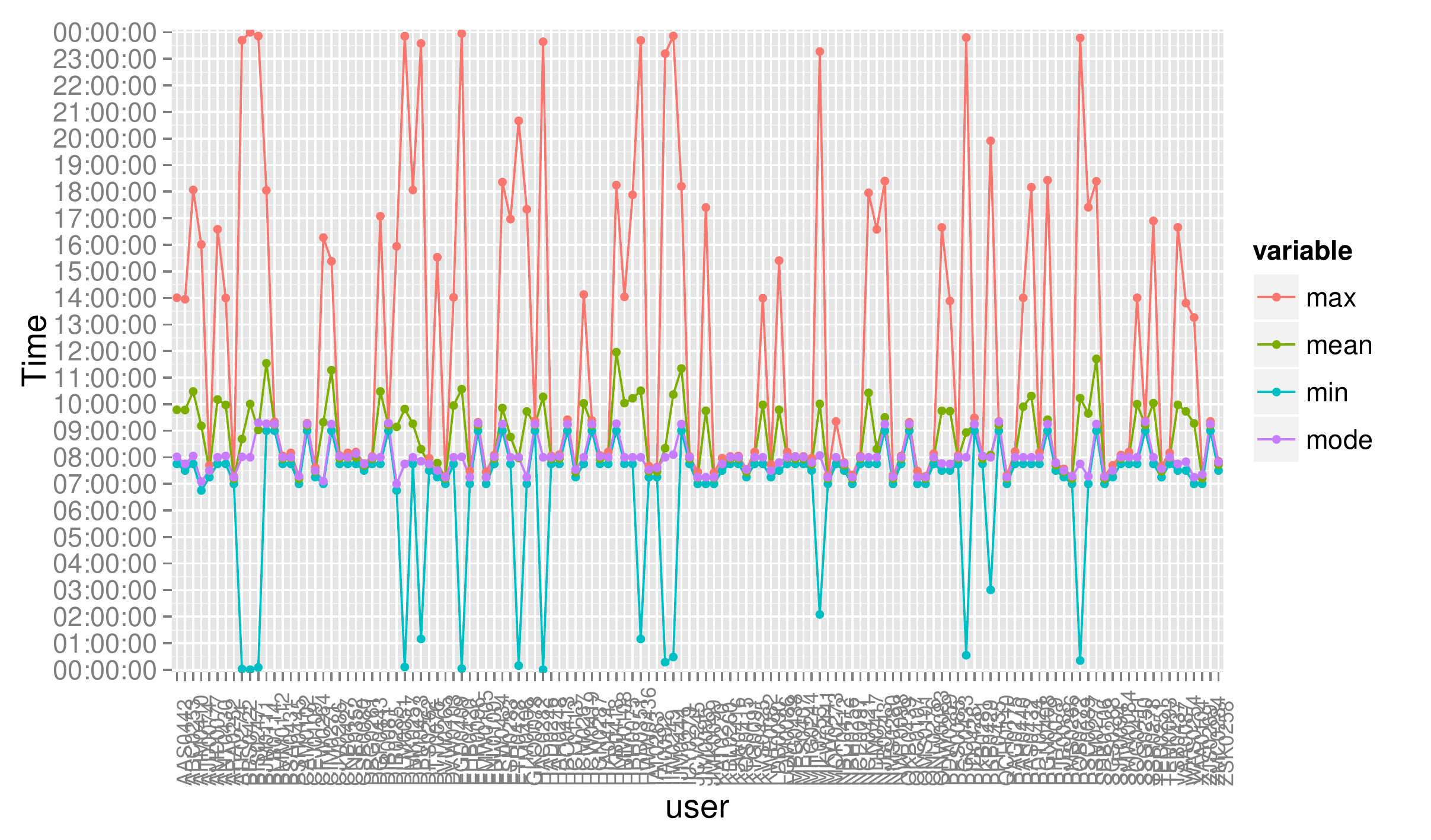} 
		\label{Logon_Behavior}}
		\subfloat[Logoff behavior]{\includegraphics[scale=0.30]{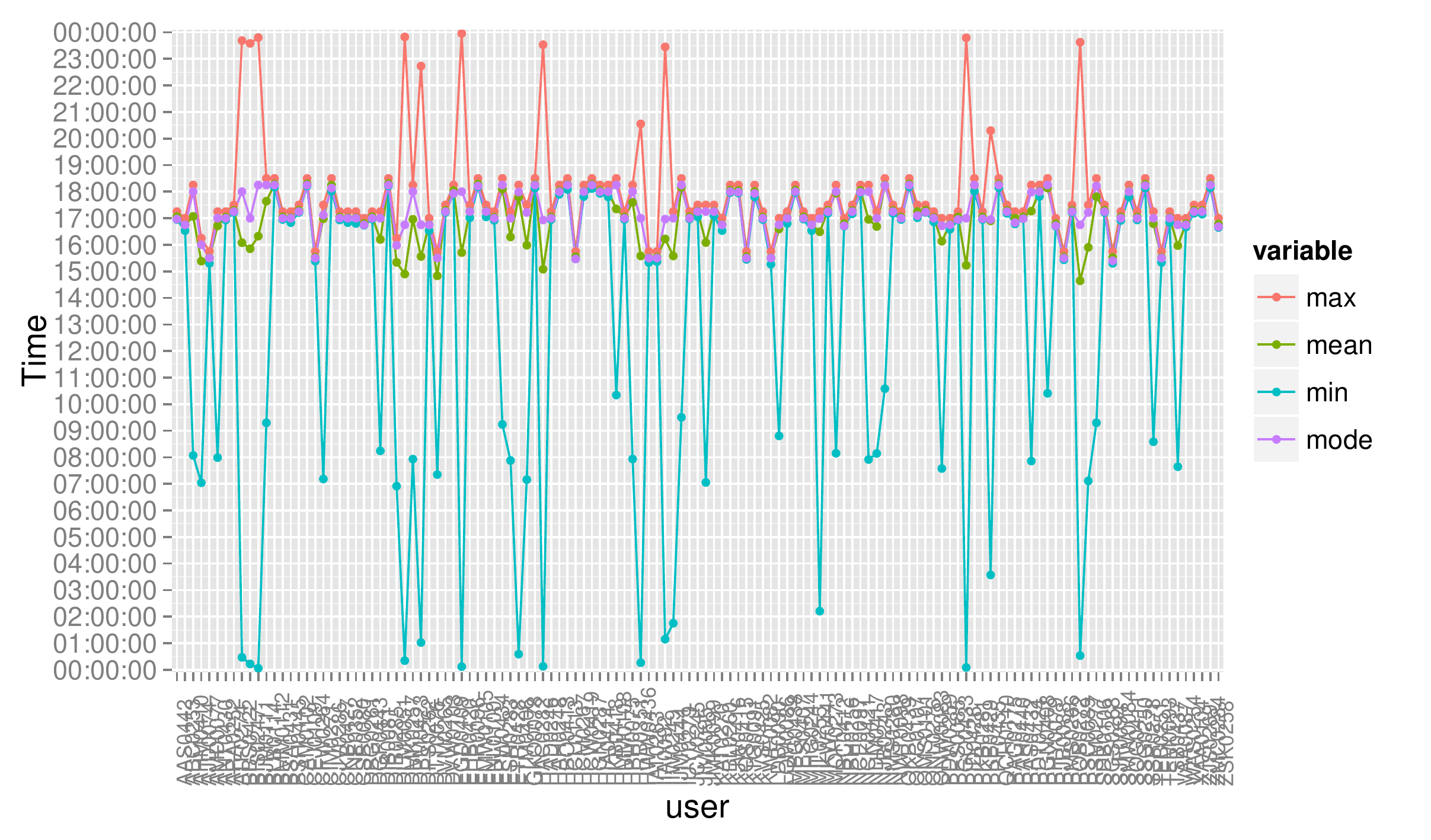}	
		\label{Logoff_Behavior}}
		\caption{Users' logon and logoff behavior}
		\label{figure 6}
		\vspace{-5pt}
\end{figure*}

\begin{figure*}[h]
		\centering
		\subfloat[USB connect]{\includegraphics[scale=0.29]{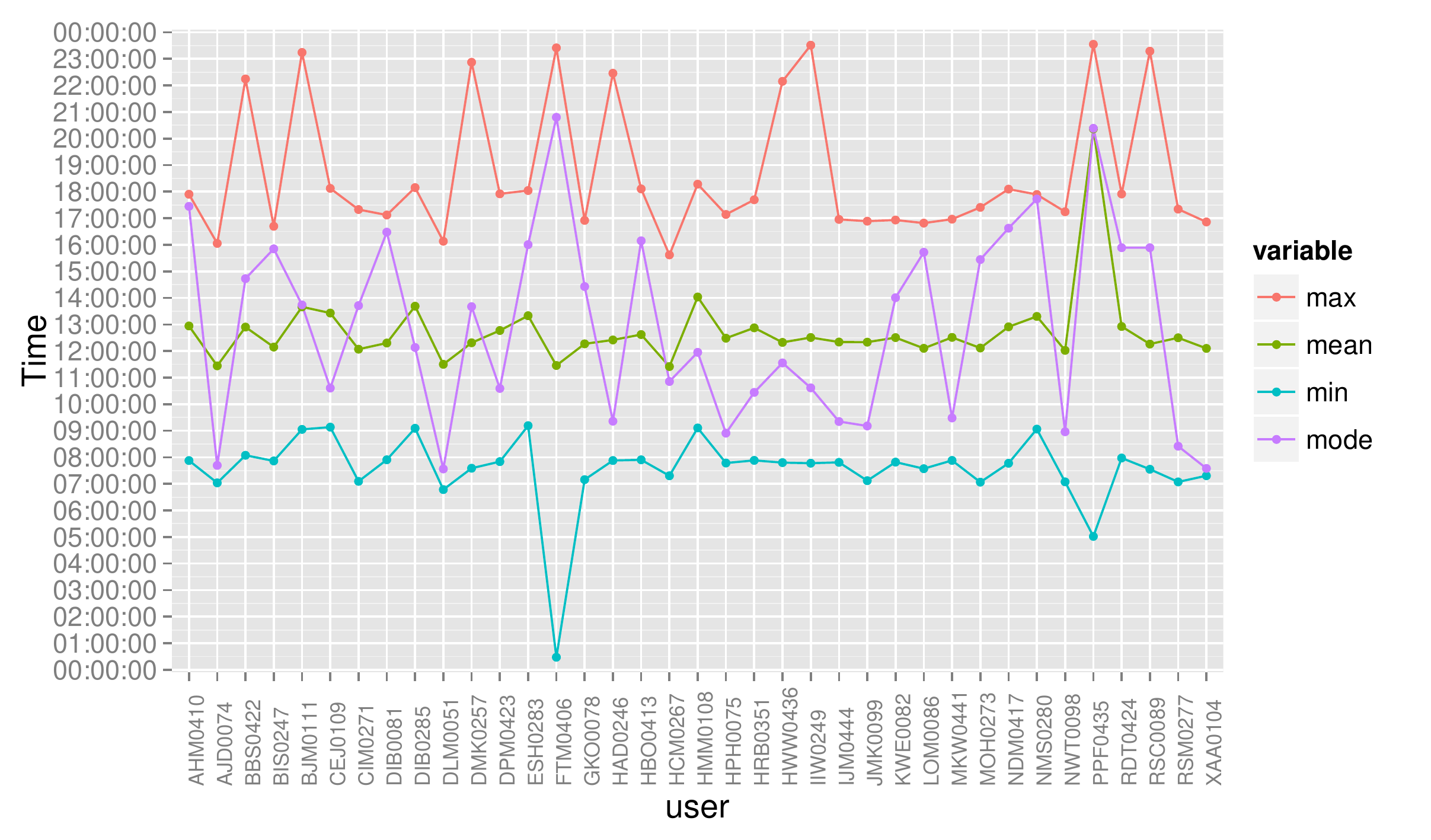} 
		\label{USB Connect}}
		\subfloat[USB disconnect]{\includegraphics[scale=0.29]{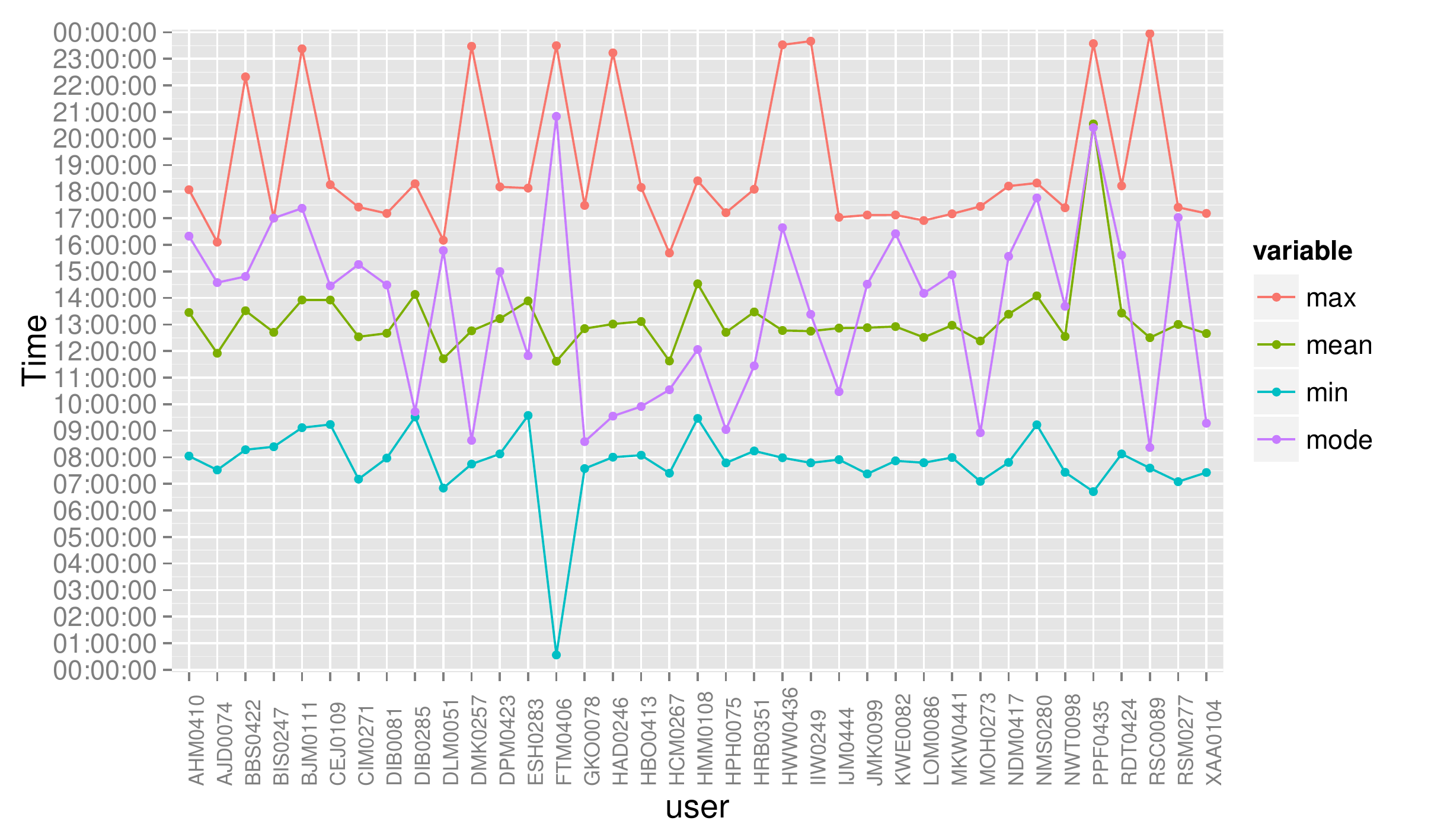}	
		\label{USB Disconnect}}
		\hfill
		\subfloat[USB file transfer statistics]{\includegraphics[scale=0.29]{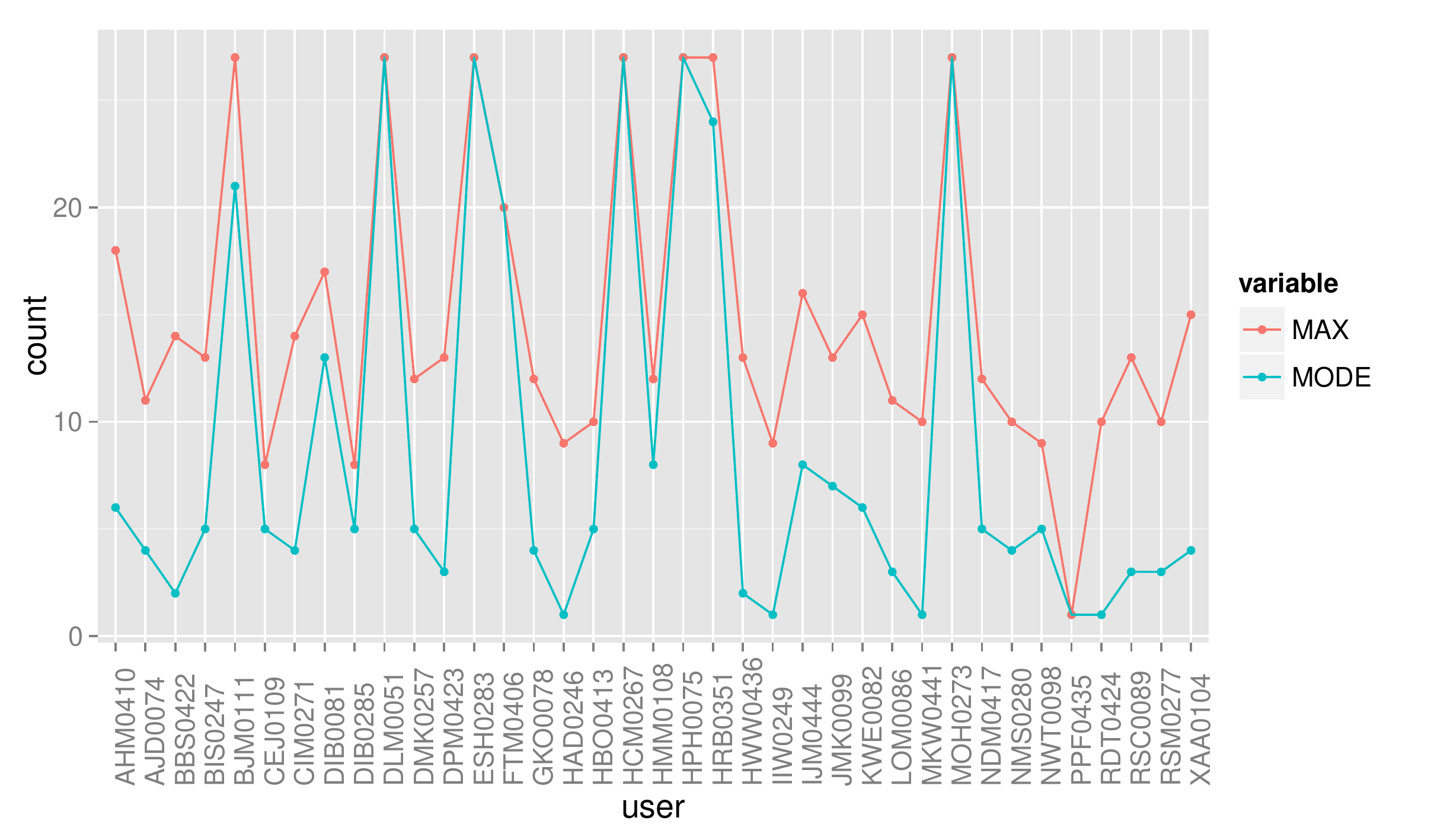} 
		\label{USB File Count}}
		\subfloat[Difference between maximum and mode number of USB file transfers]{\includegraphics[scale=0.29]{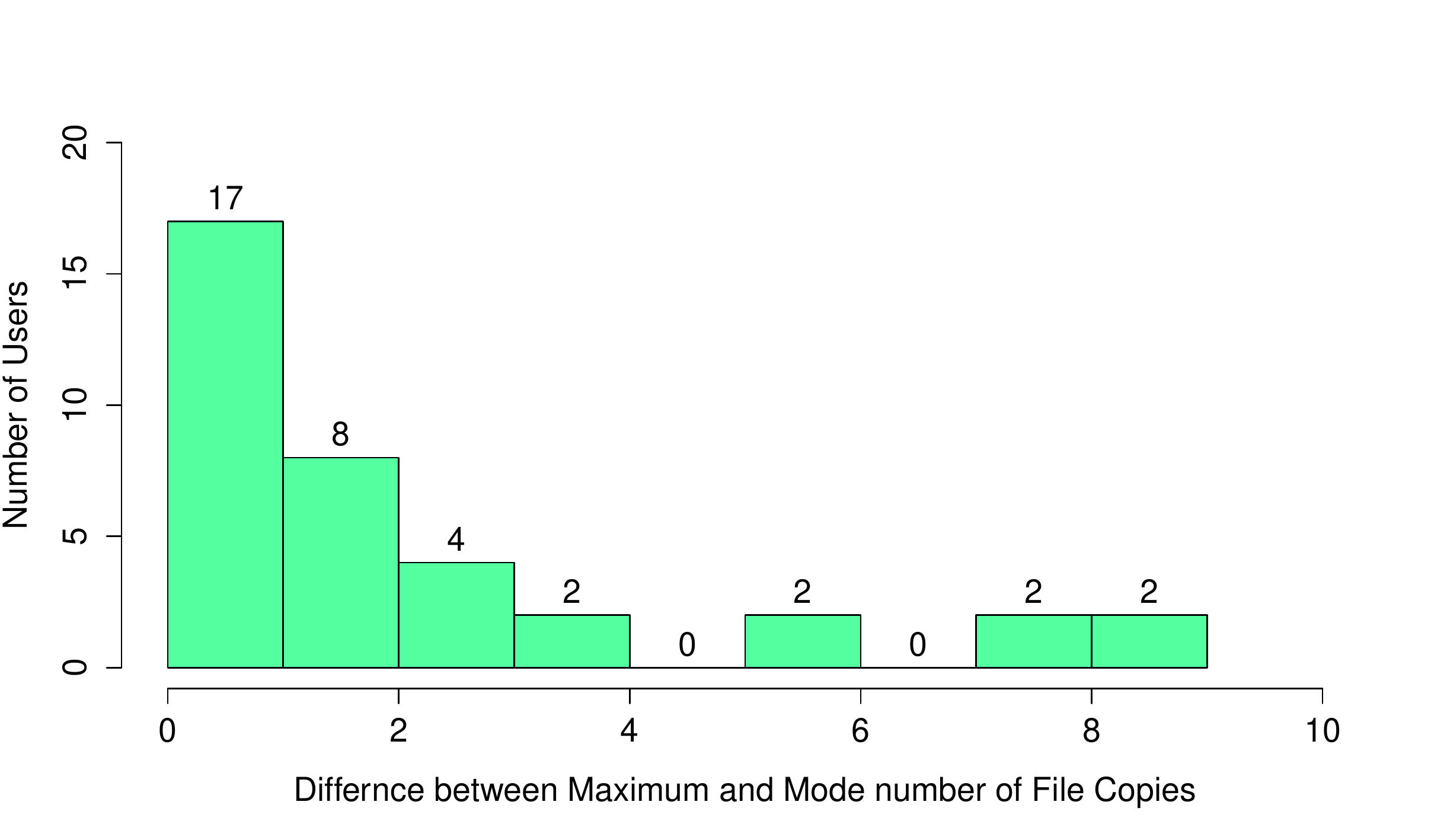}
		\label{USB Histogram}}
		\caption{Removable media usage behavior}
		\label{figure 7}
		\vspace{-5pt}
\end{figure*}

\subsubsection{Removable media usage}
Figure 7 is an illustration of users' removable media usage statistics. Similar to logon/logoff analysis, time dependencies of removable media usage has also been investigated. Figure 7(a) and (b) shows the maximum, minimum, mean and mode times for USB connect and disconnect events respectively. One important factor noticed through this analysis is only a $20\%\ (37/129)$ of employees from the selected designation used USB file copies, which can not be considered as a typical behavioral pattern among the chosen group. In this case also events which occur during regular office hours can be regarded as normal while events happened after hours can be either suspicious or work related. Also, we have to be vigilant about large file copying during after hours as well as normal business hours. This property can be yield by analysis of the time gap between consecutive connect and disconnect events, which we have not computed in this exercise. Figure 7(c) and (d) demonstrate the variation of users daily number of file accesses. To identify suspicious file copies we have considered only the maximum and mode of the number of file copies per day by an individual. If the difference between the maximum and the mode of the number of file copies is unusual, it can be considered as a suspicious file download. By looking at the histogram, it is clear that the above difference is less than four for the majority of employees while a few of the users deviated from this pattern.\\
%\subsubsection{Web Access Patterns}

\smallskip\noindent
\textbf{Web Access Patterns.} For the completeness of this work, we have selected a single parameter, based on individual users' web access patterns. Distribution of the unique number of URLs accessed by individual users is illustrated in Figure 8. This results also illustrate few outliers from rest of the group, which can be directed for further investigations. We will be exploring the means of integration of our previous work \cite{Anagi} to get more input parameters based on web access patterns in the continuation of this work.

\begin{figure}[h]
\centering
\includegraphics[scale=0.30]{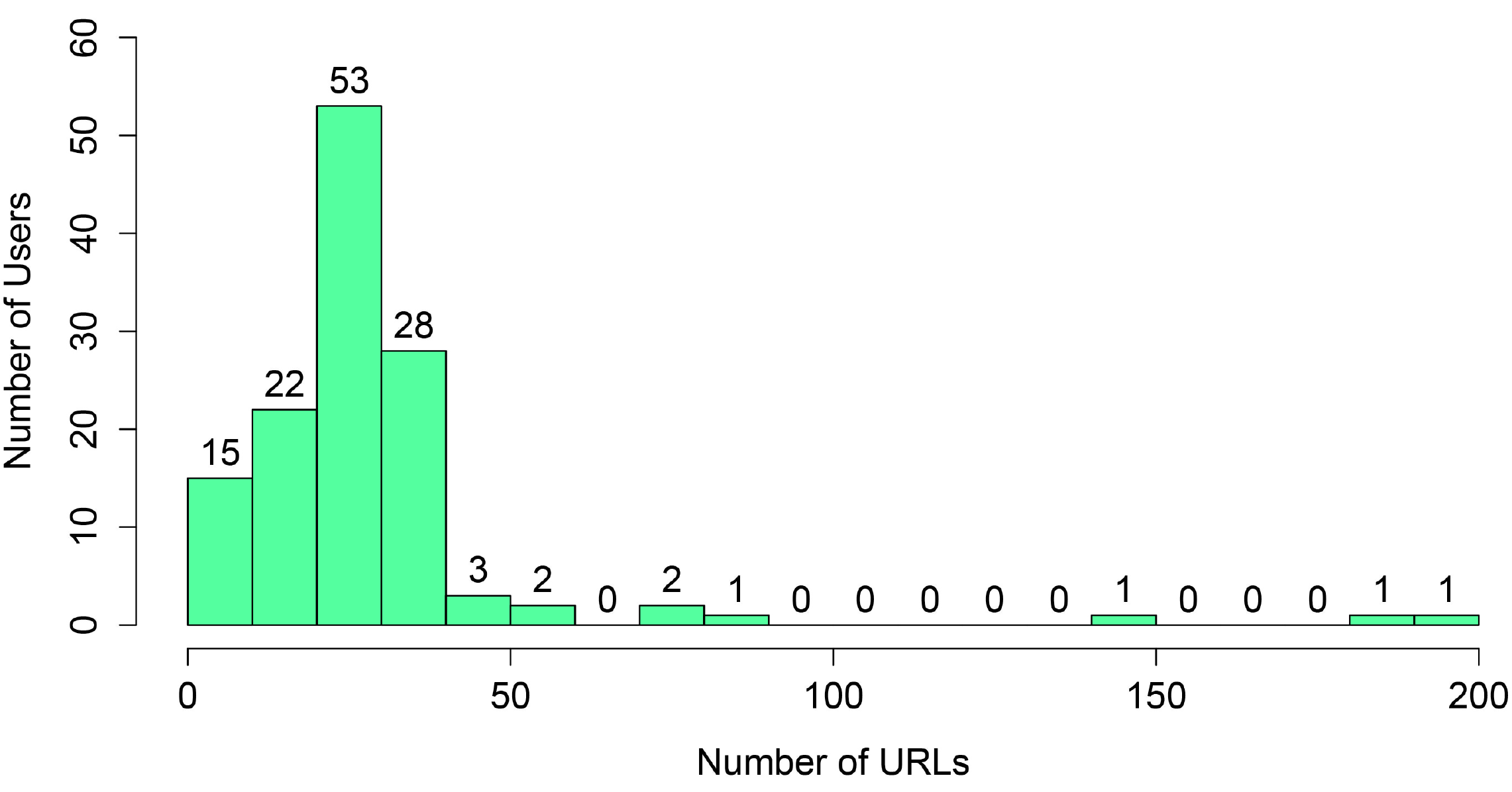}
\vspace{-5pt}
\caption{Distribution of unique URLs accessed by users}
\label{figure 8}
\vspace{-5pt}
\end{figure}

\subsection{Anomaly Detection}
Figure 9(a) illustrates the anomaly score distribution of the user base for the different order of subgraphs and the combination of all subgraph properties. Anomaly score distribution clearly indicates major two types of users based on the above-proposed subgraph properties. The users in the tail of the distribution have small anomaly score values, and they do not change with the order of subgraph. However, the other set of users who are in the main segment of the distribution have higher anomaly scores and vary based on the order of subgraph. Based on the Isolation Forest algorithm, users with anomalous scores very close to 1 can be considered as definite anomalies while the instances with anomaly scores much smaller than 0.5 are safe to consider as typical cases. 

\subsection{Parameter Dependency} 
\begin{figure}[h]
		\centering
		\subfloat[Sub graph properties]{\includegraphics[width=0.8\linewidth]{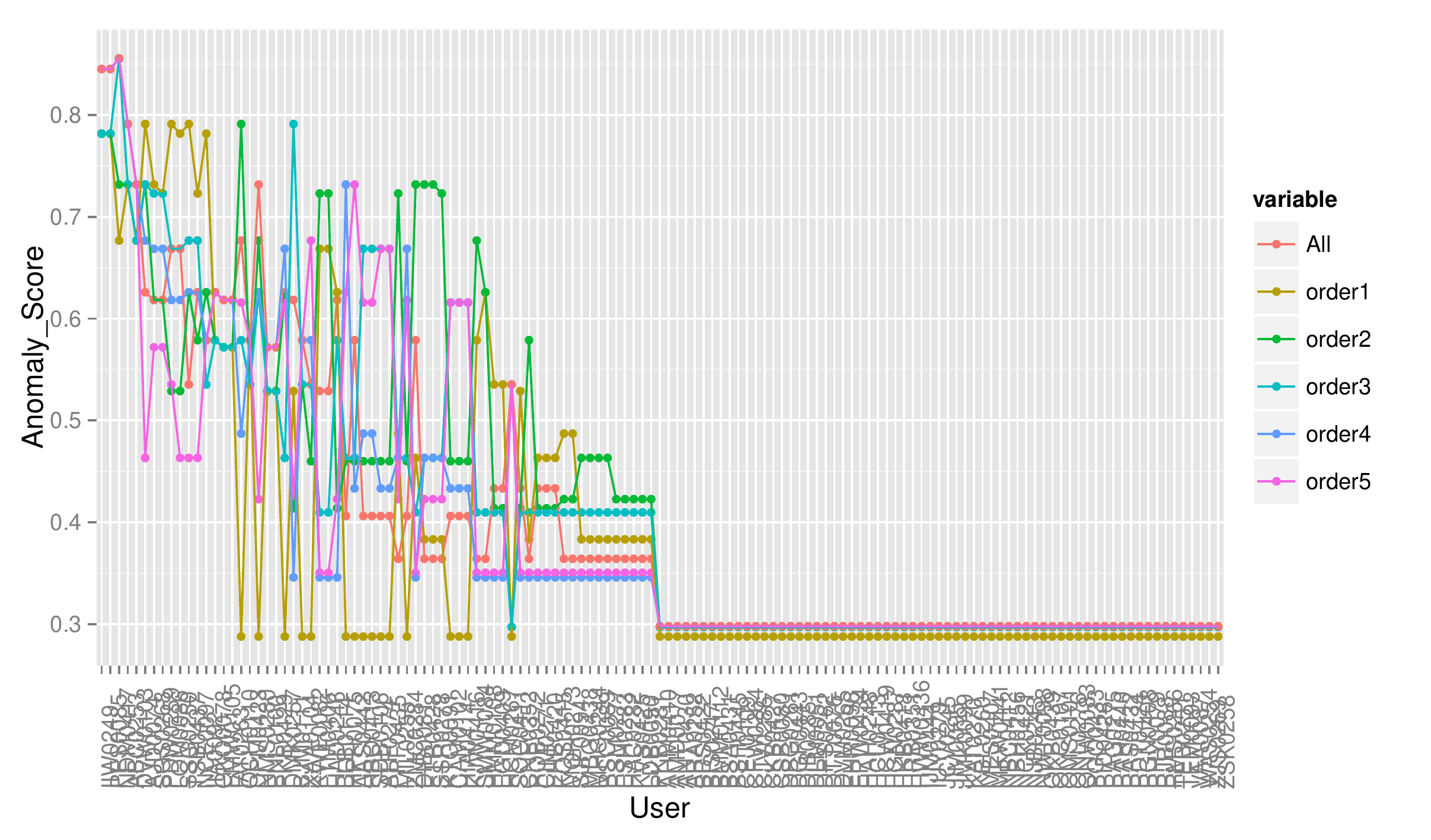}
		\label{Sub Graph Properties}}
		\hfil
		\subfloat[All properties]{\includegraphics[width=0.8\textwidth]{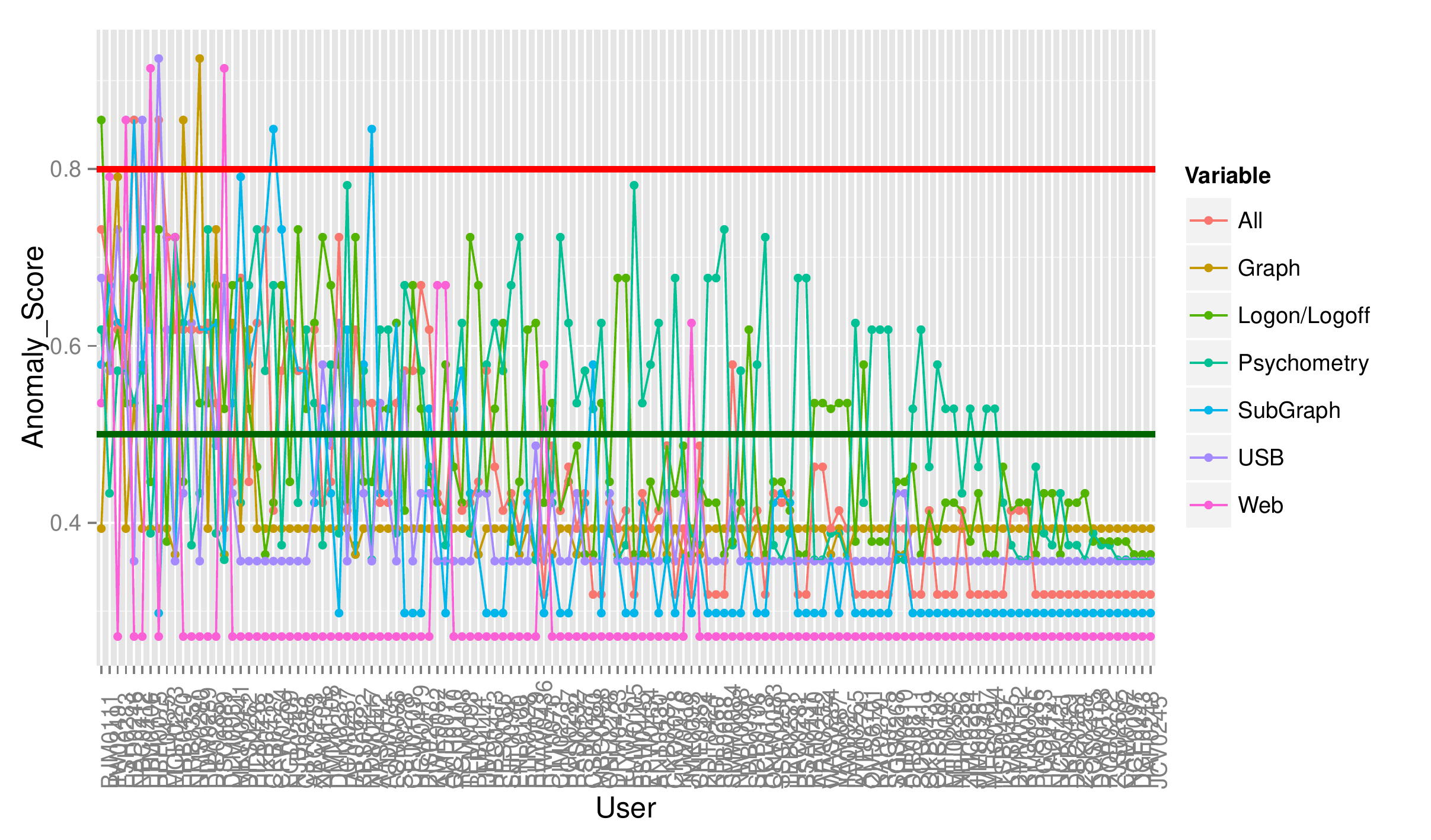}
		\label{All Properties}}
		\caption{Anomaly score distribution}
		\label{figure 9}
		\vspace{-5pt}
\end{figure}

Figure 9(b) is an indication of how the anomaly scores are distributed on different input parameters chosen in this analysis. The dense graph indicated few points above the ``Red" color horizontal line which is equivalent to an anomaly score of 0.8. Users belong to those points can be considered as anomalous users. To get a better understanding of results illustrated in Figure 9(b), the distribution of users with respect to anomaly scores computed based on all identified parameters is shown as a histogram in Figure 10. We find that the majority of users have anomaly scores in the interval $[0.4,0.7]$, which can be considered to be normal, while a minority of users have anomaly score values above 0.7 which can be regarded as suspicious. The single outlier with anomaly score value higher than 0.8 can be tagged for further investigations.

In the case of insider threat detection and prevention, the priority is on the isolation of suspicious users from the rest of workforce. Since it was the major intention of this work, we have used the following technique to validate above results by computing the percentage of suspicious users based on the different parameters identified. To perform that, the calculated anomaly scores are mapped into a binary vector $(0,1)$ based on a predefined threshold value for each parameter. The threshold value for each parameter is selected as the $(maximum \ value - 0.1)$, keeping a margin of $10\%$ as in most of the experimental cases. In the particular case of anomaly scores of $\approx 0.5$ for the entire dataset, we will exclude that parameter in the validation process as that parameter do not contribute much in finding suspicious activities. For all the other cases anomaly scores are mapped as described above. We propose the following two methods to check parameter dependencies as discussed below. 

\textbf{\textit{Case I }}:
In this approach percentage of suspicious users are calculated considering the predicted number of anomalous parameters at a time from $1\ to\ 6$. Calculated percentages are summarized in Table 4. Calculated percentage values reveal more than $79\%$ of users can be considered to have normal behavior, while the others have suspicious behavior. Also noted there are no users who are suspicious when considered more than three parameters together. The complex nature of insider threat problem can govern this type of results.

\textbf{\textit{Case II }}:
In this method, we proposed looking at the all possible combinations of parameters from the six main categories selected for this analysis. After considering the results of above Case I, we could not expect a significant change in results. However, we believe this would be a possible approach in identification of parameter dependencies.
\begin{table*}[h]
\caption{Statistics on parameter dependency}
\centering
\begin{tabular}{|P{0.22\linewidth}|p{0.16\linewidth}|p{0.045\linewidth}|p{0.045\linewidth}|p{0.045\linewidth}|p{0.045\linewidth}|p{0.045\linewidth}|p{0.045\linewidth}|p{0.045\linewidth}|}
\hline 
\multirow{2}{*}{Functional unit } & \multirow{2}{*}{Department} & \multicolumn{7}{c|}{Percentage of Users Corresponds to Number of Anomalous Parameters (\%)}\\
\cline{3-9} 
 &  & 0 & 1 & 2 & 3 & 4 & 5 & 6\\
\hline Research And Engineering & Engineering & 83.72 & 16.28 & 0.00 & 0.00 & 0.00 & 0.00 & 0.00\\
\hline  Research And Engineering & Software Management & 82.18	&  15.84 & 0.99 & 0.99 & 0.00 & 0.00 & 0.00 \\
\hline  Research And Engineering & Research & 79.21 & 17.82 & 2.97 & 0.00 & 0.00 & 0.00 & 0.00 \\
\hline 
\end{tabular}
\vspace{-5pt}
\end{table*}

\begin{figure}[h]
\centering
\includegraphics[scale=0.25]{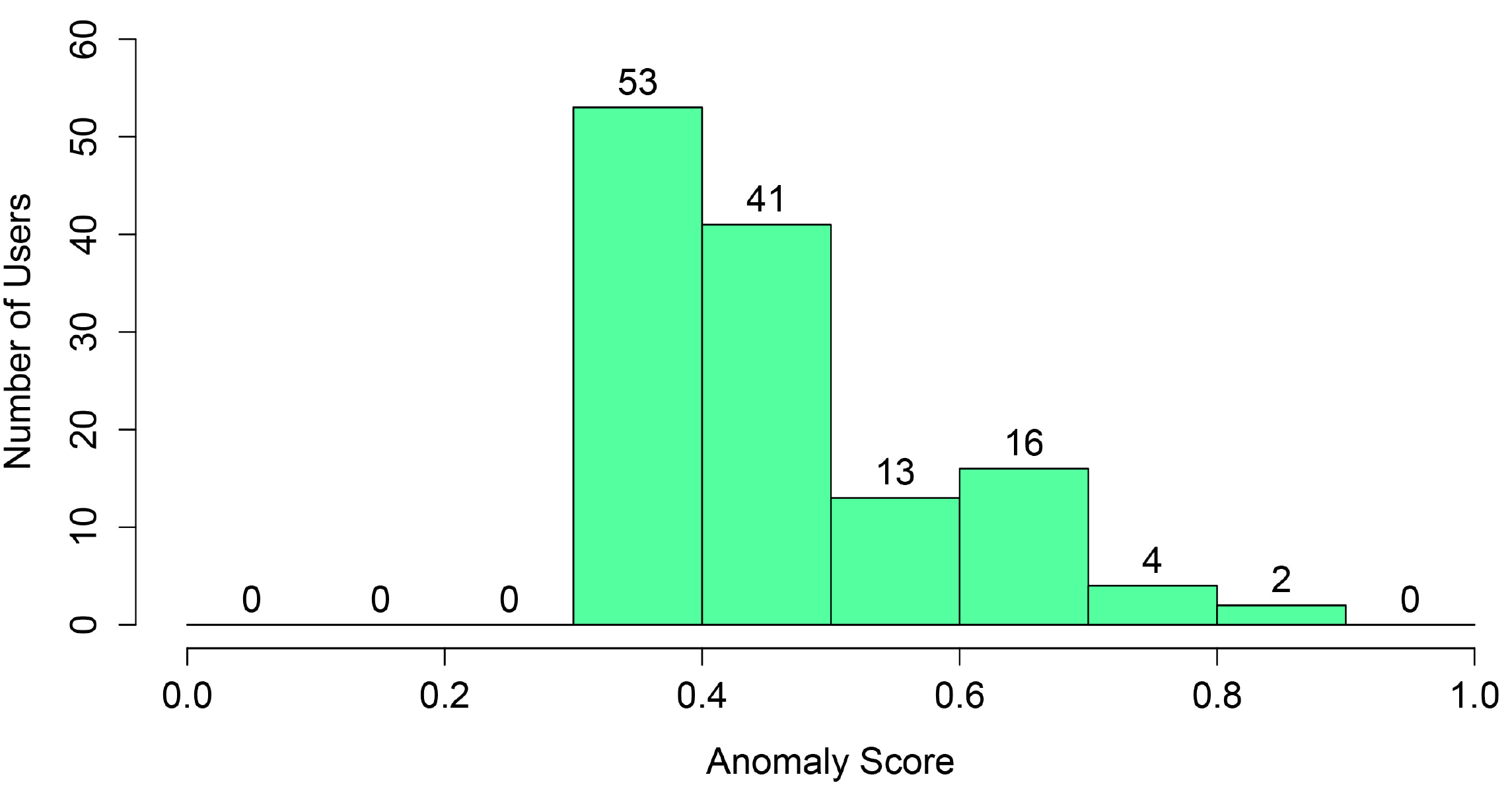}
\vspace{-5pt}
\caption{Anomaly score distribution of users}
\label{figure 10}
\vspace{-5pt}
\end{figure}

\section{Conclusions and Future Work}
In this paper, we have introduced a framework based on graphical and anomaly detection approaches for identifying potential malicious insiders. This model generates anomaly scores based on different input parameters for each user. Considering the nature of insider attacks a user can be deemed to be suspicious even if a single parameter has been found to be suspicious. We have adopted graph, subgraph properties and statistical methods in generating input parameters for the anomaly detection algorithm through multi-domain real world information. Empirical results reveal the importance of selected properties in combating this patient and smart attack. We also found that more than $79\%$ of users with common behavioral patterns while the rest of the group shows suspicious behavior based on different parameters. Users belong to the minority group can be tagged and directed for further investigation.

In the continuation of this work, we will focus on integrating as many as possible input parameters to improve the effectiveness of the proposed framework. It would include social network data which can be considered as a good source of online behavior and other statistical inputs from email and instant messaging communications. We believe this framework would be much more useful if we can include data from content analysis of other sources such as websites access by users, emails sent/received by users and file transfers to removable media. We will focus on integrating several other graphical parameters, few other statistical parameters such as time gap between USB insert/remove actions in expanding this framework. Also, we will explore the possibilities of integration of similarity based clustering mechanism which we have introduced in our previous work for web access pattern analysis \cite{Anagi}, with the proposed model. Another focus is on extending the analysis for the entire dataset and validate results with the $2$ insider threat cases simulated in the dataset. The other main aspect we are looking towards is the use of temporal properties in graph analysis to incorporate time factor instead of statistical analysis. Finally, the main goal of this work, as well as our previous and future work is to formulate an efficient, scalable and automated insider threat detection and prediction framework.

\section*{Acknowledgments}

The authors were supported by  Australian Research Council (ARC) linkage grants, LP110200321 and LP140100698. In addition, Anagi Gamachchi is supported by an ATN (Australian Technological Network) Industrial Doctoral Training Center (IDTC) PhD scholarship.

\bibliographystyle{IEEEtranS}

\bibliography{ref}

\end{document}